\documentclass[aps,prd,nofootinbib,preprint,showkeys,floatfix]{revtex4}
\usepackage{graphicx}
\usepackage{amsmath,amssymb}
\usepackage{slashed}
\usepackage{url}
\usepackage{multirow}

\def\rpv{$\slashed{R}_p \:$}
\def\mrpv{\slashed{R}_p \:}

\newcommand{\AddrAHEP}{
  {\it AHEP Group, Instituto de F\'{\i}sica Corpuscular --
    C.S.I.C./Universitat de Val{\`e}ncia \\
    Edificio de Institutos de Paterna, Apartado 22085,
  E--46071 Val{\`e}ncia, Spain}}

\newcommand{\AddrOrsay}{%
Laboratoire de Physique Th\'eorique, CNRS -- UMR 8627, \\Universit\'e de Paris-Sud 11, F-91405 Orsay Cedex, France
}

\begin{document}

\preprint{IFIC/12-52,LPT/12-80}

\title{Neutrino masses from R-parity violation with a $Z_3$ symmetry}

\author{E. Peinado} \email{epeinado@ific.uv.es}
\affiliation{\AddrAHEP}
\author{A. Vicente}\email{avelino.vicente@th.u-psud.fr}
\affiliation{\AddrOrsay}
\keywords{supersymmetry; neutrino masses and mixing}

\pacs{14.60.Pq, 12.60.Jv, 14.80.Cp}

\begin{abstract}
We consider a supersymmetric model where the neutrino mass matrix
arises from bilinear and trilinear R-parity violation, both restricted
by a $Z_3$ flavor symmetry. Assuming flavor blind soft supersymmetry
(SUSY) breaking conditions, corrected at low energies due to running
effects, we obtain a neutrino mass matrix in agreement with
oscillation data. In particular, a large $\theta_{13}$ angle can be
easily accommodated. We also find a correlation between the reactor
and atmospheric mixing angles. This leads in some scenarios to a clear
deviation from $\theta_{23} = \pi/4$. The Lightest Supersymmetric
Particle (LSP) decay, dominated by the trilinear couplings, provides a
direct way to test the model at colliders.

\end{abstract}

\maketitle

\section{Introduction}

Supersymmetry (SUSY) is one of the most popular extensions of the
Standard Model (SM) \cite{Martin:1997ns,Nilles:1983ge}. The main
advantage with respect to non-SUSY models is the technical solution
that it provides to the hierarchy problem
\cite{Gildener:1976ai,Veltman:1980mj,Sakai:1981gr,Witten:1981nf}.  In
addition, other theoretical and phenomenological issues can be
addressed in the context of supersymmetric models. For example,
questions such as the radiative origin of the electroweak symmetry
breaking \cite{Ibanez:1982fr,Ellis:1983bp,AlvarezGaume:1983gj} and the
unification of the gauge couplings at high energies
\cite{Langacker:1990jh,Ellis:1990wk,Amaldi:1991cn,Langacker:1991an,Giunti:1991ta}.

However, despite these appealing theoretical motivations, no
experimental evidence of supersymmetry has been found so far at the
Large Hadron Collider (LHC) \cite{Chatrchyan:2011zy,Aad:2011ib}. This
might be telling us that we are adopting a wrong search strategy. In
fact, most searches are based on the assumption of conserved R-parity
\cite{Farrar:1978xj}, thus including a lower cut on the amount of
missing transverse energy. This should encourage the search for
non-minimal supersymmetric scenarios with a departure from the usual
signatures, such as those with R-parity violation \cite{Hall:1983id}.

In the Minimal Supersymmetric Standard Model (MSSM) there are
renormalizable terms, perfectly allowed under all symmetries of the
theory, that break either baryon or lepton number
\cite{Hall:1983id}. These new interactions, that involve SM particles
and their superpartners, are highly constrained by the non-observation
of baryon or lepton number violating processes\footnote{Many detailed
  studies regarding bounds on R-parity violating couplings have been
  written. These are obtained from baryon or lepton number violating
  processes as well as from several (unobserved) flavor violating
  processes. See for example
  \cite{Choudhury:1996ia,deGouvea:2000cf,Barbier:2004ez,Dreiner:2006gu,Kao:2009fg,Dreiner:2010ye,Dreiner:2012mn,Dreiner:2012mx}
  and references therein.}.  In fact, if baryon and lepton number
breaking interactions were present at the same time, they would induce
proton decay. For these reasons, these dangerous terms are usually
forbidden by an \emph{ad-hoc} additional symmetry, such as R-parity.

However, in order to stabilize the proton it is sufficient to forbid
one of these terms, namely the baryon number violating one. In this
case, the lepton number violating interactions would contribute to the
generation of Majorana neutrino masses
\cite{Hall:1983id,Ross:1984yg}, an open issue not
addressed in the MSSM, while problems due to proton decay are
evaded. This well motivated framework is one of the few scenarios
where the mechanism behind neutrino mass generation can be directly
tested in accelerator experiments, see for example
\cite{Mukhopadhyaya:1998xj,Hirsch:2000ef,Porod:2000hv,Hirsch:2002ys,Bartl:2003uq,Thomas:2011kt,Hanussek:2012eh,deCampos:2012pf}.

Furthermore, the presence of R-parity violating couplings opens up
good perspectives for a richer phenomenology. Although the smallness
of the R-parity violating couplings implies that the standard decay
chains are not affected, the decay of the LSP changes dramatically the
final signatures at the LHC \cite{Dreiner:2012wm}. In fact, the LSP
decay reduces the amount of missing transverse energy in the final
state. Instead, one expects events with large lepton and/or jet
multiplicity. This has been recently used by different authors
\cite{Graham:2012th,Hanussek:2012eh} to relax the stringent bounds on
the squark and gluino masses, otherwise pushed to values clearly above
the TeV.

Regarding neutrino oscillation experiments, the intense activity over
the last years has led to an increasing accuracy in the determination
of the oscillation parameters. In particular, the $\theta_{13}$ mixing
angle has been finally measured, with a surprisingly large
result. Indeed, Double-Chooz~\cite{Abe:2011fz},
T2K~\cite{Hartz:2012pr}, MINOS~\cite{Adamson:2012rm},
Daya-Bay~\cite{An:2012eh} and RENO~\cite{Ahn:2012nd} have completely
ruled out the possibility of a vanishing $\theta_{13}$, consistently
pointing towards a value in the $\sin^2 2 \theta_{13} \sim
\mathcal{O}(0.1)$ ballpark. Updated global fits to all available
experimental data have also appeared recently
\cite{Tortola:2012te,Fogli:2012ua,SchwetzTalk}.

The aforementioned measurement of $\theta_{13}$ has also ruled out the
popular tri-bimaximal neutrino mixing pattern \cite{Harrison:2002er}
and led to an explosion of papers where different flavor symmetries
are used to accommodate a large $\theta_{13}$. Concerning
supersymmetric models with R-parity violation, a flavor model for
bilinear R-parity violation was introduced in
\cite{Bazzocchi:2012ve}. The requirement of an exact $A_4$ flavor
symmetry in the superpotential couplings led to the known neutrino
mixing pattern at the prize of a higher complexity of the scalar
sector of the model. Here we present another example, based on a
simple $Z_3$ symmetry, of a flavored R-parity violating model. In
addition to the bilinear term, the trilinear ones play a fundamental
r\^ole in the generation of neutrino masses. The flavor symmetry
strongly restricts the allowed terms, and only a few remain in the
model. However, it still allows for the required interplay that leads
to the observed neutrino masses and mixings.

The paper is organized as follows. In section \ref{sec:model} we
define the model and the basic assumptions followed along the
paper. In section \ref{sec:numass} we describe the origin of the
different contributions to the neutrino mass matrix and we analyze the
resulting neutrino mixing pattern in
section \ref{sec:mixing}. Finally, we briefly comment on collider
phenomenology in section \ref{sec:collider} and conclude in
section \ref{sec:conclusions}.

\section{Definition of the model}
\label{sec:model}

We consider the MSSM particle content with the following charge assignment under an additional $Z_3$ symmetry:

\begin{table}[h!]
\begin{center}
\begin{tabular}{|c|c|c|c|c|c|c|c|c|}
\hline
 & $\,\hat{L}_1\,$ & $\,\hat{L}_2\,$ & $\,\hat{L}_3\,$ & $\,\hat{E}_1\,$ & $\,\hat{E}_2\,$ & $\,\hat{E}_3\,$ & $\,\hat{H}_d\,$ & $\,\hat{H}_u\,$\\
\hline
$Z_3$ & $1$& $\omega$&$\omega^2$& $1$& $\omega^2$& $\omega$& $1$& $1$\\
\hline
\end{tabular}\caption{Charge assignment of the model. We use the standard notation $\omega = e^{i 2\pi/3}$.}\label{model}
\end{center}\end{table}

In addition, all quark superfields are singlets under $Z_3$. The most general superpotential in the leptonic sector allowed by the gauge and flavor symmetries is
\begin{equation} \label{superpotential}
\mathcal{W} = Y_e^i \hat{L}_i \hat{E}_i \hat{H}_d + \epsilon \hat{L}_1 \hat{H}_u + \lambda_{ijk} \hat{L}_i \hat{L}_j \hat{E}_k + \lambda'_{jk} \hat{L}_1 \hat{Q}_j \hat{D}_k,
\end{equation}
where the only non-zero $\lambda_{ijk}$ couplings allowed by $Z_3$ are $\lambda_{122}$, $\lambda_{133}$ and $\lambda_{231}$. The flavor symmetry also implies that only one $\epsilon$ parameter is non-zero, $\epsilon \equiv \epsilon_1$, the other two being forbidden due to the $\mu$ and $\tau$ charges under $Z_3$. We do not specify the superpotential terms without lepton fields for brevity, since they do not play any r\^ole in the following discussion. We also impose the conservation of baryon number. This forbids the superpotential term $\lambda''_{ijk} U_i D_j D_k$ and estabilizes the proton.

We allow for a soft breaking of the $Z_3$ flavor symmetry in the scalar potential, and thus general terms of the type
\begin{equation} \label{soft-pot}
V_{\text{soft}} \sim m_{L_i H_d}^2 \tilde{L}_i^* H_d + B_\epsilon^i \tilde{L}_i H_u,
\end{equation}
are considered.

Finally, in the following discussion we will make two important assumptions about the parameters of the model. First, we will assume that the $\epsilon$ parameter is a small dimensionful parameter and thus it plays a negligible r\^ole in neutrino mass generation. That will allow us to drop some contributions to the neutrino mass matrix, as required to obtain the correct neutrino mixing pattern. In the next section we will estimate how small $\epsilon$ must be for our discussion to be unaltered. Second, we will consider universal soft SUSY breaking terms at the grand unification scale, $m_{GUT}$. This implies that generation dependent couplings are exactly the same for all families and, in particular, one has $m_{L_1 H_d}^2 = m_{L_2 H_d}^2 = m_{L_3 H_d}^2$ and $B_\epsilon^1 = B_\epsilon^2 = B_\epsilon^3$ at $m_{GUT}$. However, this equality is no longer valid at the SUSY scale due to the effect of Renormalization Group Equations (RGE) running. In fact, in CMSSM-like scenarios, like the one considered here, 3rd family sfermions soft terms depart from universality at low energies due to their larger Yukawa couplings\footnote{Another possibility that takes non-universality in the soft terms as starting point is the so-called \emph{natural supersymmetry}, where the 1st and 2nd generation sfermions are much heavier than the 3rd one, see for example \cite{Baer:2012uy,Badziak:2012rf}.}. Therefore, we will simply make use of this departure from universality and consider independent 3rd family sfermion soft terms at the SUSY scale\footnote{Approximate formulas for the 3rd family soft terms are known in the literature, see \cite{Codoban:1999fp}. However, their complexity (and dependence on unknown parameters such as $\tan \beta$) does not allow for simple estimates. Therefore, we will treat the departure from universality as a free phenomenological parameter.}.

\section{Neutrino mass generation}
\label{sec:numass}

The breaking of lepton number induces non-zero Majorana masses for the neutrinos which, in this model, are generated by both bilinear and trilinear R-parity violating couplings. As a result of that, the neutrino mixing pattern requires the correct interplay between those two sources. In this section we will describe the different contributions to the neutrino mass matrix and estimate their sizes.

The interplay between different \rpv couplings is not a new idea. For example, a similar approach was followed in Ref. \cite{Bhattacharyya:2011zv}, where the combination of bilinear and trilinear R-parity violation, together with a properly chosen flavor symmetry, also led to the required structure for the neutrino mixing matrix. However, the construction of the model and the resulting neutrino mass matrix are different in our work. The use of flavor symmetries to restrict the allowed \rpv couplings has also been discussed in \cite{Banks:1995by,Ellis:1998rj,Bhattacharyya:1998vw,Kajiyama:2007pr} and, in particular, the $Z_3$ group has also been considered in \cite{Du:1994bf,Liu:1996ua}. Furthermore, many discrete symmetries have been used as a substitute for R-parity, see for example the recent \cite{Dreiner:2011ft,Dreiner:2012ae}.

Finally, before we proceed to describe the different contributions to the neutrino mass matrix we note that the charge assignment in table \ref{model} leads to a diagonal charged lepton mass matrix.

\subsection{Bilinear R-parity violation}
\label{subsec:numass1}

A precise determination of the 1-loop corrections to neutrino masses in bilinear R-parity violating supersymmetry requires the consideration of several types of diagrams. We will follow closely the results presented in the pioneer references on this topic \cite{Grossman:1997is,Grossman:1998py,Davidson:2000uc,Grossman:2003gq}.

The $Z_3$ flavor symmetry implies that only one bilinear $\epsilon$ parameter is allowed in the superpotential, $\epsilon \hat{L}_1 \hat{H}_u$. This would in principle lead to one non-zero entry in the tree-level neutrino mass matrix, $m_\nu^{11} \propto \epsilon^2$. However, we will assume that the smallness of the $\epsilon$ parameter makes this contribution negligible.

\begin{figure}[tb]
\centering
\includegraphics[width=0.7\linewidth]{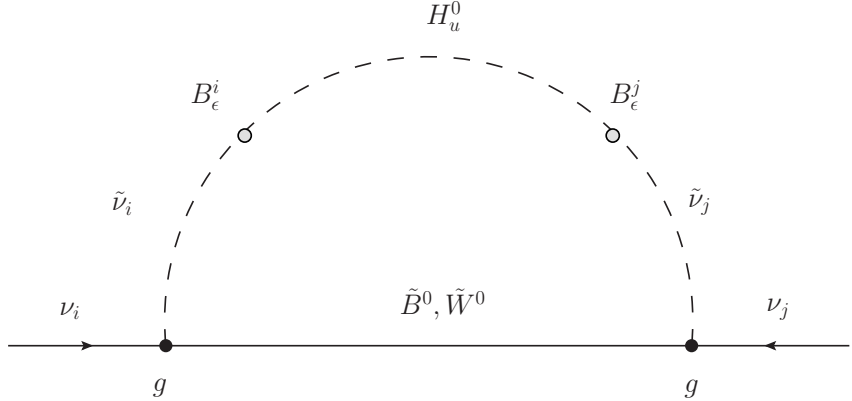}
\caption{1-loop neutrino masses from bilinear R-parity violation. $B_\epsilon$ contribution.}
\label{BRpV-loop-1}
\end{figure}

\begin{figure}[tb]
\centering
\includegraphics[width=0.7\linewidth]{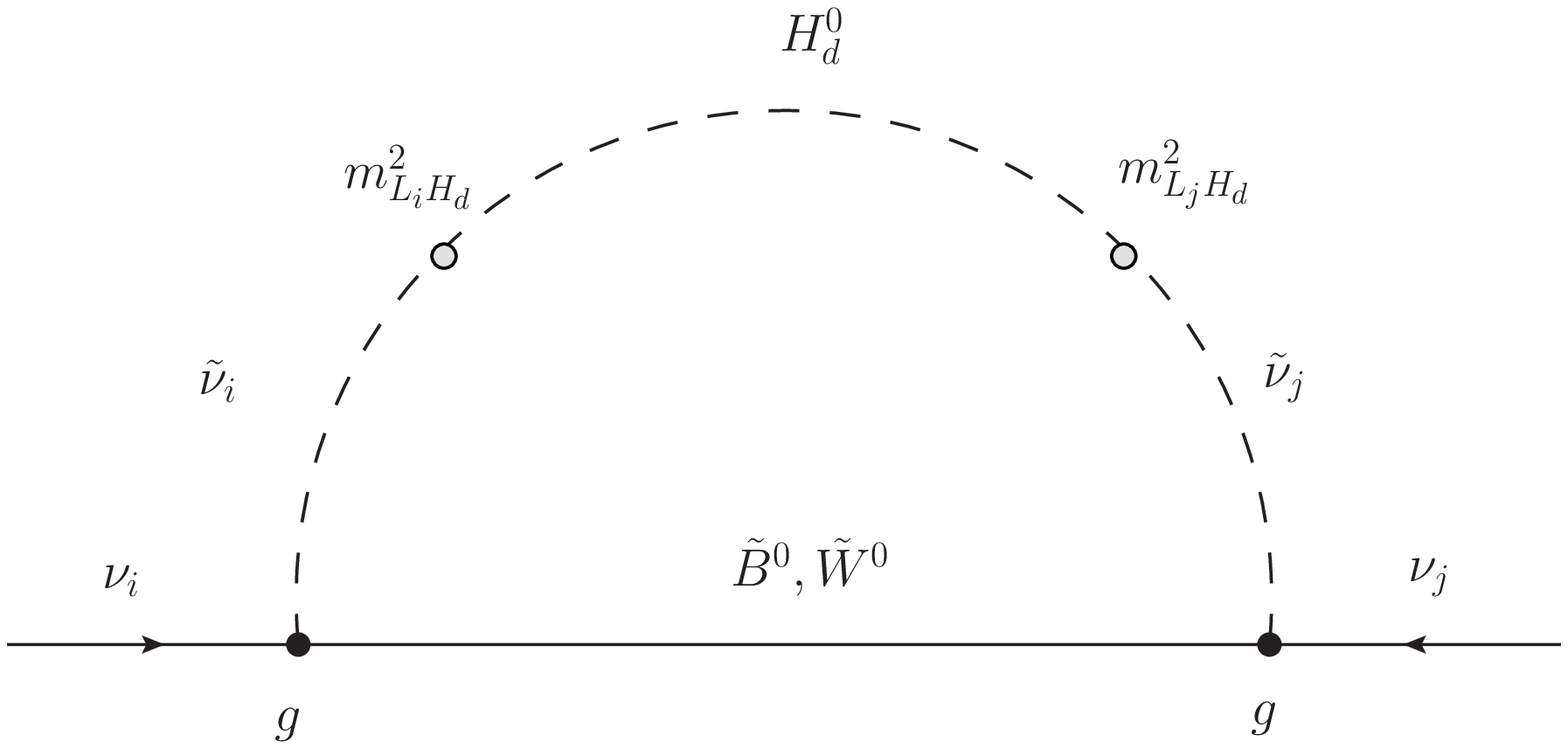}
\caption{1-loop neutrino masses from bilinear R-parity violation. $m_{L H_d}^2$ contribution.}
\label{BRpV-loop-2}
\end{figure}

Therefore, the dominant BRpV contributions to neutrino masses come from 1-loop diagrams without the $\epsilon$ coupling (see Ref. \cite{Abada:2001zh} for a similar scenario). Examples of such possibilities are presented in figures \ref{BRpV-loop-1} and \ref{BRpV-loop-2}. We note however that they require relatively large $B_\epsilon^i \, , \, m_{L_i H_d}^2$ (see below for an estimate), otherwise these 1-loop contributions would be too small to account for neutrino masses. This would generate large sneutrino vacuum expectation values (VEVs), $v_L^i$, and lead to too large tree-level contributions to neutrino masses. Therefore we are forced to impose a fine-tuning of the parameters so that the relative contributions to $v_L^i$ coming from $B_\epsilon^i$ and $m_{L_i H_d}^2$ cancel each other. This fine-tuning relation can be easily obtained from the potential in equation \eqref{soft-pot},
\begin{equation} \label{fine-tuning}
m_{L_i H_d}^2 v_d + B_\epsilon^i v_u \simeq 0,
\end{equation}
which implies that the approximate relation $m_{L_i H_d}^2 \simeq \tan \beta \, B_\epsilon^i$ must hold.

We proceed now to estimate the size of the different diagrams. If the scale of lepton number violation in the soft terms is denoted as $B_\epsilon^i \, , \, m_{L_i H_d}^2 \sim m_{\mrpv}^2$, one obtains\footnote{In fact, this definition of $m_{\mrpv}^2$ actually corresponds to $m_{L_i H_d}^2$. That is why equation \eqref{BRpV-estimate} does not include a $\tan^2 \beta$ enhancement, see \cite{Davidson:2000uc, Grossman:2003gq}, since that factor is absorbed in $m_{\mrpv}^2$.}
\begin{equation} \label{BRpV-estimate}
m_{\nu}^{\text{BRpV}} \text{(1-loop)} \sim \frac{g^2}{16 \pi^2} \, \frac{m_{\mrpv}^4}{m_{\text{SUSY}}^3}.
\end{equation}
For $m_{\text{SUSY}} = 100$ GeV one finds that $m_{\mrpv} \sim 1$ GeV leads to $m_{\nu}^{\text{BRpV}} \sim 0.1$ eV, in the correct range. On the other hand, the tree-level contribution to the neutrino mass matrix can be roughly estimated to be $m_{\nu}^{\text{BRpV}} \text{(tree)} \sim \epsilon^2 / m_{\text{SUSY}}$, and thus for $m_{\text{SUSY}} = 100$ GeV one needs $\epsilon \gtrsim 100$ KeV to be competitive with the 1-loop contributions. We will therefore assume in the following that $\epsilon$ is below that scale.

It is important to notice that these contributions require a non-vanishing mass splitting between the real and imaginary components of the sneutrinos, $\Delta m_{\tilde{\nu}}^2$, as discussed in \cite{Grossman:1997is} and previously deduced from general considerations in \cite{Hirsch:1997vz}.

Concerning the flavor structure of these 1-loop contributions, note that by assumption the soft SUSY breaking potential is flavor universal for the 1st and 2nd generations, with a (little) departure in the 3rd generation soft terms. Therefore, we obtain the sum of two terms, a democratic structure plus a deviation from universality
\begin{equation}\label{mass1}
m_{\nu}^{\text{BRpV}}= a \left(
\begin{array}{ccc}
1 & 1 & 1 \\
1 & 1 & 1 \\
1 & 1 & 1
\end{array}
\right) + d \left(
\begin{array}{ccc}
0 & 0 & 1 \\
0 & 0 & 1 \\
1 & 1 & 2
\end{array}
\right).
\end{equation}
Here $a$ is given by the estimate in equation \eqref{BRpV-estimate}. The dimensionful coefficient $d$ follows from a similar expression, where the corresponding soft terms $m_{\mrpv}^2$ have been properly replaced by $m_{\mrpv}^2+\delta m^2$. The mass squared parameter $\delta m^2$ is the result of the deviation from universality, expected to be sizable for the 3rd generation. Terms of order $(\delta m^2)^2$ have been neglected in equation \eqref{mass1}.

Before we move on to the discussion of the contributions from trilinear R-parity violating couplings, we would like to emphasize that the deviation from universality provided by the second term in Eq. \eqref{mass1} is phenomenologically required. As we will see below, this is the only term that breaks $\mu-\tau$ invariance in the neutrino sector. Its absence would imply a vanishing reactor mixing angle and maximal atmospheric mixing, see for instance \cite{Caravaglios:2005gw}. This would be in contradiction with recent data, that clearly indicates $\theta_{13} \ne 0$. Therefore, one can interpret the departure from universality in the soft terms, quite general in supersymmetric models, as the origin of the non-zero reactor angle in our model\footnote{As already explained, the $\mu-\tau$ invariance can also be broken \emph{by hand} by introducing a small difference between the soft SUSY breaking terms $B_\epsilon^2$ and $B_\epsilon^3$ (or, analogously, $m_{L_2 H_d}^2$ and $m_{L_3 H_d}^2$). Similarly, one can relax the conditions given by the $Z_3$ symmetry by allowing the existence of $T_\lambda$ couplings whose corresponding superpotential terms are forbidden. For example, $T_{\lambda_{233}}$ would induce a high order breaking of the $\mu-\tau$ invariance of the neutrino mass matrix. However, we consider the breaking by the Yukawa couplings, transferred to the soft SUSY breaking terms by RGE running, the simplest solution.}.

\subsection{Trilinear R-parity violation}
\label{subsec:numass2}

\begin{figure}[tb]
\centering
\includegraphics[width=0.7\linewidth]{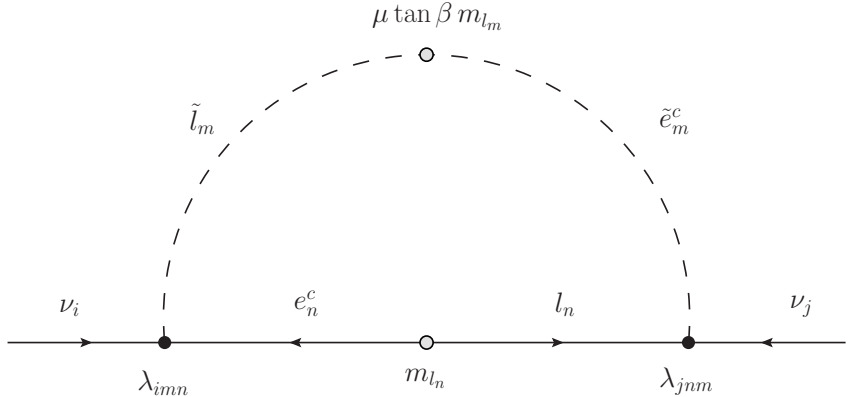}
\caption{1-loop neutrino masses from trilinear R-parity violation.}
\label{TRpV-loop}
\end{figure}

As previously explained, the flavor symmetry implies that the only non-zero $\lambda_{ijk}$ couplings are $\lambda_{122}$, $\lambda_{133}$ and $\lambda_{231}$. Therefore, the usual 1-loop diagrams in trilinear R-parity violation, see figure \ref{TRpV-loop}, can only fill the $11$ and $23-32$ entries of the neutrino mass matrix, leading to
\begin{equation}\label{mass2}
m_{\nu}^{\text{TRpV}}=\left(
\begin{array}{ccc}
b & 0 & 0 \\
0 & 0 & c \\
0 & c & 0
\end{array}
\right).
\end{equation}

Similarly, the $\lambda'$ couplings can also contribute to neutrino masses, but only to the $11$ element of the matrix, here denoted by $b$. We can now estimate the size of this family of 1-loop contributions. For example, in the case of the $c$ element in the previous matrix one has
\begin{equation} \label{TRpV-estimate}
\left( m_{\nu}^{\text{TRpV}} \right)_{23} = c \sim \frac{1}{16 \pi^2} \, \lambda_{231}^2 \mu \tan \beta \, \frac{m_e m_\tau}{m_{\text{SUSY}}}.
\end{equation}
The element $b$ has a similar generic expression, although more couplings are involved (and for the $\lambda'$ diagrams quark masses appear instead). Equation \eqref{TRpV-estimate} shows that $\mathcal{O}(0.1$ eV) contributions can be obtained for $\mu \sim m_{\text{SUSY}} = 100$ GeV and $\tan \beta = 10$ if $\lambda_{231} \sim 0.05$. This value is close to the experimental limit, but it can be easily lowered by using $\mu > m_{\text{SUSY}}$ or larger $\tan \beta$. See table \ref{bounds} and refs. \cite{Barbier:2004ez,Kao:2009fg} for more details.

Summing up the two contributions one obtains the texture
\begin{equation}\label{masstotal}
m_{\nu} = m_{\nu}^{\text{BRpV}} + m_{\nu}^{\text{TRpV}}=\left(
\begin{array}{ccc}
a+b & a & a+d \\
a & a & a+c+d \\
a+d & a+c+d & a+2d
\end{array}
\right).
\end{equation}

In the following section we will show how the mass matrix in Eq. \eqref{masstotal} can accommodate the observed pattern of neutrino mixing. However, we already observe that the coefficient $d$, which can be traced back to the departure from universality in the soft SUSY breaking terms, is the only source of $\mu-\tau$ invariance breaking. In fact, in the limit $d = 0$, one recovers an exactly $\mu-\tau$ symmetric neutrino mass matrix that implies $\theta_{13}$ and maximal atmospheric angle\footnote{We remind the reader that we are working in a basis where the charged lepton sector is diagonal.}. In conclusion, $d \ne 0$ is required. This will induce a non-zero reactor angle and a deviation from maximal atmospheric mixing. Since the same parameter is at the root of both deviations, we expect a tight correlation between $\theta_{13}$ and $\theta_{23}$.

\section{Non-zero $\theta_{13}$ and deviations from maximal atmospheric mixing}
\label{sec:mixing}

The mass matrix in Eq. \eqref{masstotal} has only four parameters. Therefore, since it must be able to accommodate the pattern of neutrino mixing measured in oscillation experiments, some connections between the parameters must appear. Although we cannot give definite predictions for the oscillation parameters, clear correlations between them exist. Similarly, the absolute scale of neutrino masses, determinant for neutrinoless double beta decay experiments, is linked to the other parameters. This allows to test the model with neutrino phenomenology.

In our analysis we assumed real parameters\footnote{This assumption will be lifted below when the general CP violating case is considered.}, so the number of free parameters in the neutrino sector is $4$, namely, $a$, $b$, $c$, and $d$ in Eq. (\ref{masstotal}), to explain $6$ observables, namely two squared mass differences $\Delta m_{atm}^2$ and $\Delta m_{sol}^2$, the neutrinoless double beta decay effective mass $m_{ee}$, and the three mixing angles. Therefore we have two predictions: the neutrinoless double beta decay effective mass and a correlation between the reactor mixing angle and the atmospheric mixing angle.

We have performed a detailed scan over the parameter space fixing $\Delta m_{atm}^2$, $\Delta m_{sol}^2$, $\sin^2 \theta_{12}$ and $\sin^2 \theta_{23}$ within the $3 \sigma$ range given in \cite{Tortola:2012te}. Large deviations from the measured value of $\theta_{13}$ have been allowed in order to clearly see potential correlations between this parameter and the other mixing angles. Correct fits\footnote{In the following, we will use the expressions \emph{valid fit} or \emph{correct fit} to denote a set of values for the parameters $a$, $b$, $c$ and $d$ that leads to $\Delta m_{atm}^2$, $\Delta m_{sol}^2$, $\sin^2 \theta_{12}$ and $\sin^2 \theta_{23}$ within the $3 \sigma$ range given in \cite{Tortola:2012te}.} have been found for both, normal and inverted neutrino spectra, without a particular preference. However, the results clearly favor a quasi-degenerate spectrum with large neutrino masses.

\begin{figure}[tb]
\centering
\includegraphics[width=0.49\linewidth]{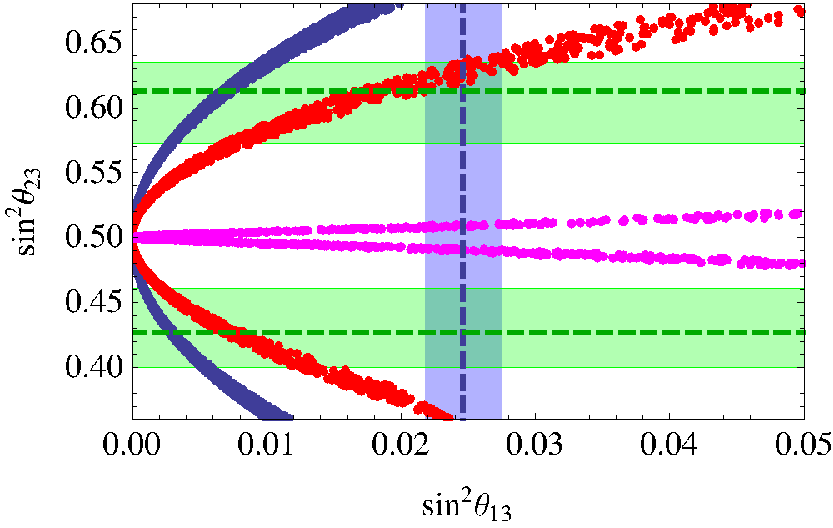}
\includegraphics[width=0.49\linewidth]{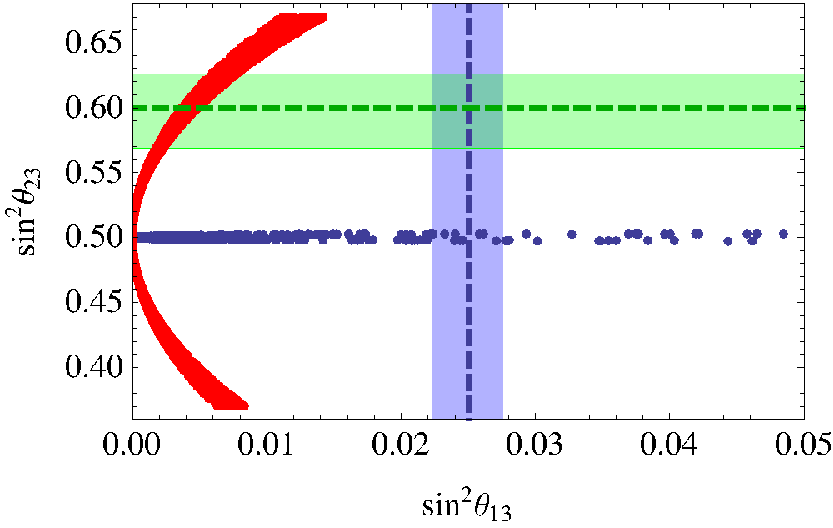}
\caption{$\sin^2 \theta_{23}$ as a function of $\sin^2 \theta_{13}$
  for normal (left-side) and inverted (right-side) neutrino
  spectra. The numerical scan is based on the results of the global
  fit of Ref. \cite{Tortola:2012te}. The dashed lines represent the
  best-fit values for $\sin^2 \theta_{13}$ and $\sin^2 \theta_{23}$,
  whereas the shaded regions correspond to $3 \sigma$ (in case of
  $\sin^2 \theta_{13}$) and $1 \sigma$ (in case of $\sin^2
  \theta_{23}$) allowed regions. Different colors correspond to
  different CP branches: $\eta_1$ (blue), $\eta_2$ (red) and $\eta_3$
  (purple).}
\label{correlation}
\end{figure}

The mass matrix in Eq. \eqref{masstotal} would be $\mu-\tau$ symmetric in the absence of the $d$ parameter, thus leading to $\sin^2 \theta_{23} = 1/2$ and $\sin^2 \theta_{13} = 0$ with a free $\theta_{12}$. Interestingly, deviations from this scheme typically affect simultaneously $\theta_{13}$, that departs from zero, and $\theta_{23}$, that departs from maximality. This generic prediction, present in many flavor models (see for instance~\cite{reactoratm}, is especially relevant in our model, since it is caused by a single parameter, $d$, which leads to a tight correlation. Our numerical results confirm this expectation, see Fig. \ref{correlation}, where $\sin^2 \theta_{23}$ is shown as a function of $\sin^2 \theta_{13}$ for normal (left-side) and inverted (right-side) neutrino spectra. The atmospheric mixing angle is restricted to be within the $3 \sigma$ allowed region found in \cite{Tortola:2012te}, whereas the vertical shaded band represents the $3 \sigma$ allowed region for $\sin^2 \theta_{13}$, as given by the same analysis.

Figure \ref{correlation} deserves some comments. In addition to the aforementioned correlation, one clearly observes distinct regions in the $\theta_{13}-\theta_{23}$ plane. These regions, or \emph{branches}, correspond to different cases of intrinsic CP charges of the neutrinos, $\eta$ \cite{Schechter:1981hw}. There are four possible cases: $\eta_1 = (-,+,+)$, $\eta_2 = (+,-,+)$, $\eta_3 = (+,+,-)$ and $\eta_4 = (+,+,+)$. Other choices can be reduced to one of these by means of an unphysical global sign. In Fig. \ref{correlation}, different colors are given for these CP cases: blue for $\eta_1$, red for $\eta_2$ and purple for $\eta_3$. No valid fits were found for the $\eta_4$ case. Similarly, for inverted hierarchy we could only find correct points for cases $\eta_1$ and $\eta_2$.

For normal hierarchy, the $\eta_1$ case leads to very large deviations in $\theta_{23}$ that clearly departs from maximality even for small values for $\theta_{13}$. In fact, from our results in Fig. \ref{correlation} one can conclude that the $\eta_1$ branch is ruled out by oscillation data since it cannot accommodate $\theta_{13}$ and $\theta_{23}$ simultaneously. In contrast, cases $\eta_2$ and $\eta_3$ can reproduce the measured mixing angles, although with different predictions. On the one hand, $\eta_2$ leads to big deviations in $\theta_{23}$, which gives a region in excellent agreement with oscillation data. On the other hand, $\eta_3$ leads to small deviations, at most $\Delta \sin^2 \theta_{23} \sim 0.2$, in agreement at $3\sigma$. For inverted hierarchy the results are slightly different. First, it is now $\eta_2$ the one that is ruled out due to the too large deviations induced in $\theta_{23}$. And second, for $\eta_1$ one finds very small departures from maximal atmospheric mixing, at most $\Delta \sin^2 \theta_{23} \sim 0.03$.

\begin{table}[tb]
\centering
\begin{tabular}{c c c c}
\hline
Analysis & \multicolumn{2}{c}{Best-fit $\pm 1 \sigma$} & $3 \sigma$ range \\
\hline
\multirow{2}{*}{Forero et al. \cite{Tortola:2012te}} & $\; 0.427^{+0.034}_{-0.027} \;$ & $\; 0.613^{+0.022}_{-0.04} \;$ & $\; 0.36 - 0.68 \;$ \\
& \multicolumn{2}{c}{$\; 0.600^{+0.026}_{-0.031} \;$} & $\; 0.37 - 0.67 \;$\\
\multirow{2}{*}{Fogli et al. \cite{Fogli:2012ua}} & \multicolumn{2}{c}{$\; 0.386^{+0.024}_{-0.021} \;$} & $\; 0.331 - 0.637 \;$ \\
& \multicolumn{2}{c}{$\; 0.392^{+0.039}_{-0.022} \;$} & $\; 0.335 - 0.663 \;$ \\
Gonz\'alez-Garc\'ia et al. \cite{SchwetzTalk} & $\; 0.41^{+0.037}_{-0.025} \;$ & $\; 0.59^{+0.021}_{-0.022} \;$ & $\; 0.34 - 0.67 \;$ \\
\hline
\end{tabular}
\caption{Comparison between the results of Refs. \cite{Tortola:2012te,Fogli:2012ua,SchwetzTalk} regarding $\sin^2 \theta_{23}$. In case of Forero et al. \cite{Tortola:2012te} and Fogli et al. \cite{Fogli:2012ua} the upper row corresponds to normal hierarchy and the lower row to inverse hierarchy. We show the two approximately equivalent best-fit regions found in the analysis by Forero et al. \cite{Tortola:2012te} and Gonz\'alez-Garc\'ia et al. \cite{SchwetzTalk}.}
\label{tabTheta23}
\end{table}

\begin{figure}[tb]
\centering
\includegraphics[width=0.49\linewidth]{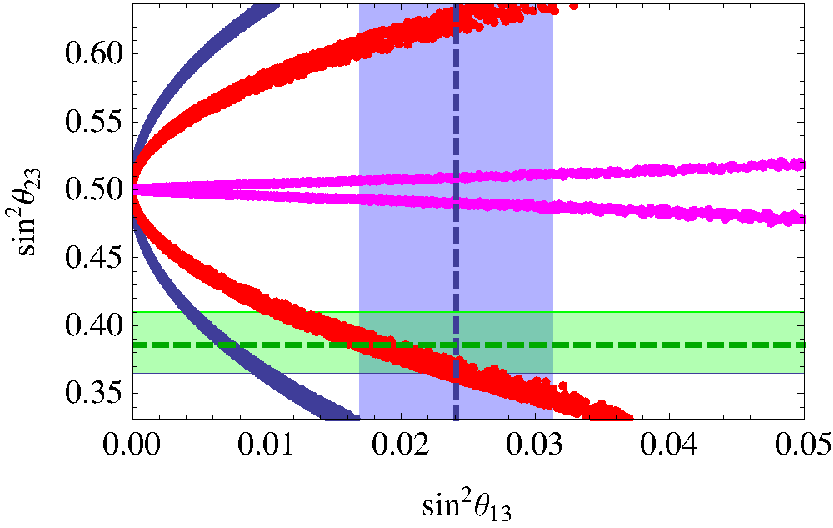}
\includegraphics[width=0.49\linewidth]{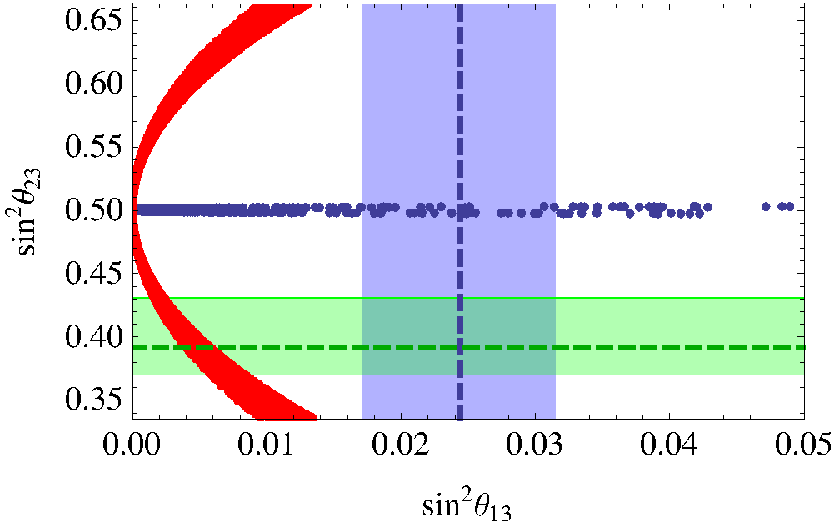}
\caption{$\sin^2 \theta_{23}$ as a function of $\sin^2 \theta_{13}$
  for normal (left-side) and inverted (right-side) neutrino
  spectra. The numerical scan is based on the results of the global
  fit of Ref. \cite{Fogli:2012ua}. The dashed lines represent the
  best-fit values for $\sin^2 \theta_{13}$ and $\sin^2 \theta_{23}$,
  whereas the shaded regions correspond to $3 \sigma$ (in case of
  $\sin^2 \theta_{13}$) and $1 \sigma$ (in case of $\sin^2
  \theta_{23}$) allowed regions. Different colors correspond to
  different CP branches: $\eta_1$ (blue), $\eta_2$ (red) and $\eta_3$
  (purple).}
\label{correlation-2}
\end{figure}

So far we have based our discussion on the results of the global fit of Ref. \cite{Tortola:2012te}. In their analysis, the authors of this work found two regions with similar ($\sim 1 \sigma$) significance for the atmospheric mixing angle in the normal hierarchy case and one region, above $\sin^2 \theta_{23} = 0.5$, in the inverse hierarchy case. A similar result is found in the analysis of \cite{SchwetzTalk}. However, the analysis \cite{Fogli:2012ua} finds a slight preference for $\theta_{23} < \pi/4$. The results of this analysis show a $\, \lesssim 2 \sigma$ deviation from maximal mixing in case of normal hierarchy, with best-fit value $\sin^2 \theta_{23} = 0.386$, and a $\, \lesssim 3 \sigma$ deviation in case of inverse hierarchy, with best-fit value $\sin^2 \theta_{23} = 0.392$. A brief compilation of these results is given in Table \ref{tabTheta23}.

Due to the impact that such deviation from maximal atmospheric mixing would have on our model, because of the correlation with the reactor angle, we have repeated our numerical scans using the input parameters given by Refs. \cite{Fogli:2012ua,SchwetzTalk}. The results for the case of Ref. \cite{Fogli:2012ua} are shown in Fig. \ref{correlation-2}. Qualitatively the results do not differ very much from those presented in Fig. \ref{correlation}. Again, there is a little preference for the $\eta_2$ case and a neutrino spectrum with normal hierarchy. We do not show the analogous figures for Ref. \cite{SchwetzTalk}, since they lead to very similar conclusions.

\begin{figure}[tb]
\centering
\includegraphics[width=0.49\linewidth]{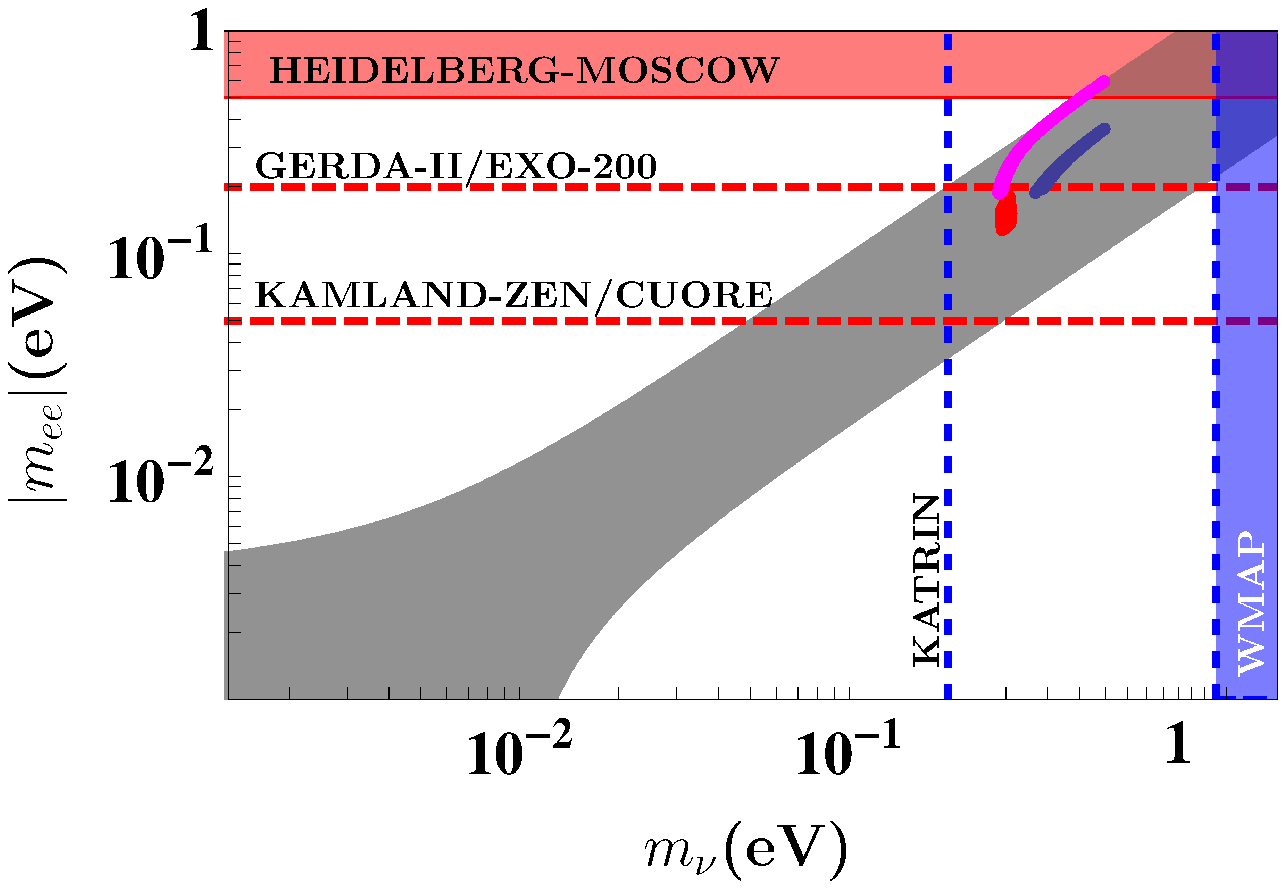}
\includegraphics[width=0.49\linewidth]{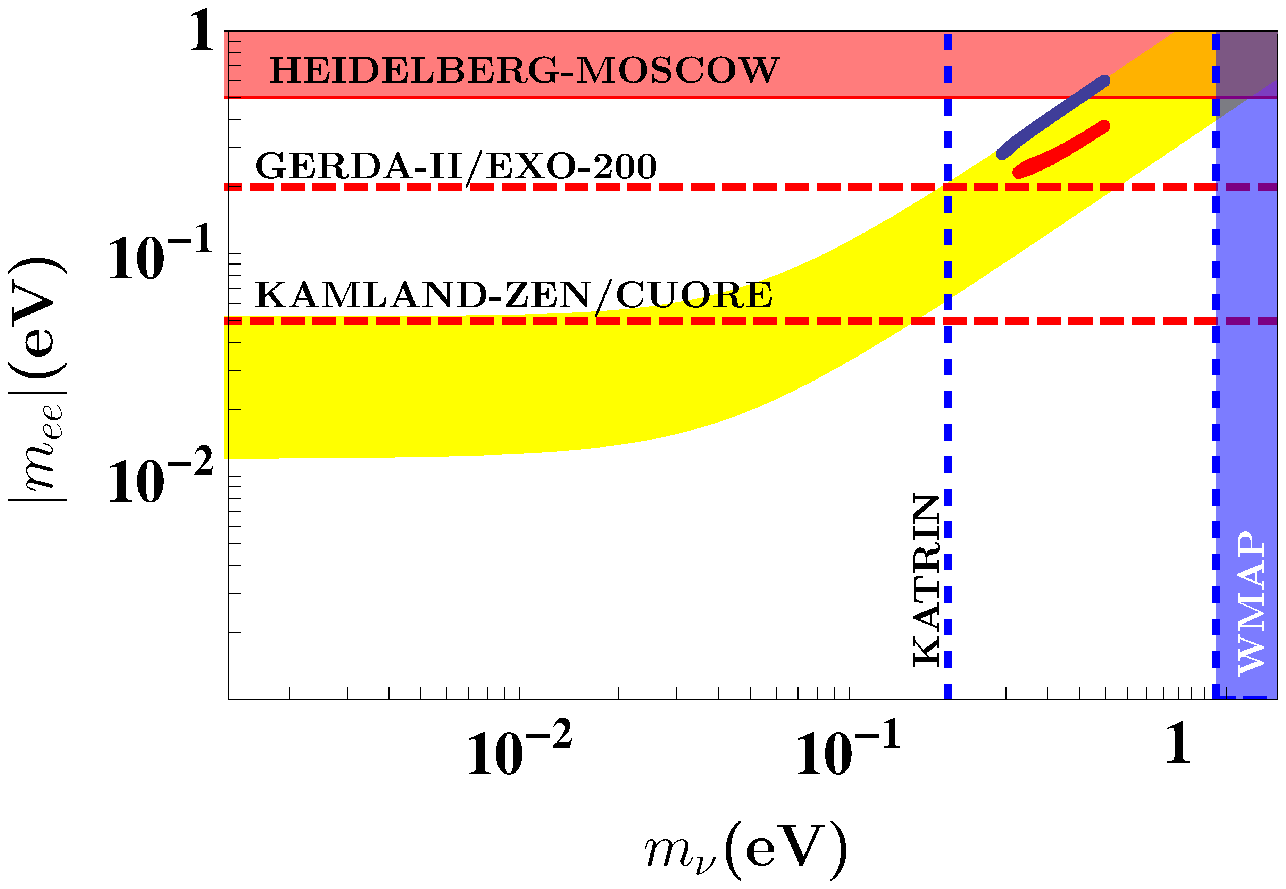}
\caption{Effective neutrinoless double beta decay parameter $m_{ee} \,$
  as a function of the lightest neutrino mass for normal (left-side)
  and inverted (right-side) neutrino spectra. The grey and yellow
  shaded regions correspond to the \emph{flavor-generic} normal and
  inverse (gray) hierarchy neutrino spectra, respectively. For
  a discussion see text.}
\label{mbb}
\end{figure}

We have also investigated the predictions that our model can make regarding neutrinoless double beta decay ($0\nu 2\beta$). The results are shown in Fig. \ref{mbb}, where the effective neutrinoless double beta decay parameter $m_{ee}$ is given as a function of the lightest neutrino mass. We present the results for normal hierarchy on the left-side and the results for inverse hierarchy on the right-side. The shaded bands (in grey and yellow) correspond to the well-known \emph{flavor-generic} predictions, much more spread than the predictions given by our model. In fact, one can easily see that our model is restricted to lie on a very small portion of the plane. This is due to the fact that only solutions with quasi-degenerate spectra were found in our scan.

Again, different CP intrinsic charges for the neutrinos lead to different regions in Fig. \ref{mbb}. The color code is as in Fig. \ref{correlation}. Future experimental sensitivities are also included in this figure, as well as the region excluded by the Heidelberg-Moscow collaboration \cite{KlapdorKleingrothaus:2000sn}. Since our model favors quasi-degenerate spectra, most of the points have large values for the lightest neutrino mass and the effective $m_{ee}$ $0\nu 2\beta$ parameter. This implies that they are at the reach of experiments such as KATRIN \cite{Osipowicz:2001sq}, GERDA-II \cite{Smolnikov:2008fu} and EXO-200 \cite{Ackerman:2011gz}\footnote{Recently, the EXO-200 collaboration made public new results \cite{Auger:2012ar}, not included in Fig. \ref{mbb}, which allow to set a lower limit on the $0\nu 2\beta$ half-time in $^{136}$Xe of about $T_{1/2} > 1.6 \times 10^{25}$ yr (90 \% CL). This corresponds to $m_{ee} < 140-380$ meV, the broad range being caused by the theoretical uncertainty in the nuclear matrix element calculation. Similarly, the KamLAND-Zen collaboration also reported new limits \cite{Gando:2012jr}, although slightly less stringent.}. Only $\eta_2$ in the case of normal hierarchy predicts low values for $m_{ee}$, out of reach for GERDA-II and EXO-200, but still with sufficiently large neutrino masses so that KATRIN is able to measure them. In conclusion, both $0\nu 2\beta$ and absolute neutrino mass experiments offer good perspectives to probe our model.

Finally, let us briefly discuss the general CP violating case, where all four parameters in the neutrino mass matrix are allowed to be complex. We have generalized our numerical scan to include all CP phases, the Dirac, $\delta$, and Majorana ones, $\alpha$ and $\beta$. The main result is that the regions between branches in Figs. \ref{correlation} and \ref{correlation-2} are filled with points where at least one of the (physical) CP phases do not vanish. Therefore, the concluions drawn in the previous paragraphs should be regarded as valid only in the CP conserving case. For example, our model would not be ruled out if the $\theta_{13}$ and $\theta_{23}$ angles are precisely measured to lie outside the $\eta_{1,2,3}$ branches. That would point towards a CP violating scenario, something to be probed by experiments such as T2K \cite{Itow:2001ee} and NO$\nu$A \cite{Ayres:2004js}, see \cite{Huber:2009cw}.

\section{Some comments on collider phenomenology}
\label{sec:collider}

Since R-parity is broken the LSP is no longer stable and
decays\footnote{This implies that the usual neutralino LSP is lost as
  dark matter candidate. However, it is well-known that a relatively
  light gravitino provides a valid alternative, see for example the
  recent references
  \cite{GomezVargas:2011ph,Restrepo:2011rj,Arcadi:2011yw}.}. The
decays can go via bilinear ($\epsilon$, $B_\epsilon^i$ and $m_{L_i
  H_d}^2$) or trilinear ($\lambda_{122}$, $\lambda_{133}$,
$\lambda_{231}$ or $\lambda'_{jk}$) \rpv couplings. In the bilinear
case, only $\epsilon$ can mediate the decay at tree-level, and thus
its small size reduces the relevance of this possibility. We are thus
left with decays mediated by trilinear couplings and we conclude that
gauge-mediated LSP decays are suppressed.

By studying the size of the different $\lambda$ and $\lambda'$
couplings one could go further and find the dominant LSP decay
signatures that particular scenarios would predict. For example, a
simple estimate based on the 1-loop generated masses that come from
$\lambda_{231}$ shows that this coupling must be of order $0.05$. On
the other hand, the other two $\lambda$ parameters ($\lambda_{122}$
and $\lambda_{133}$) are strongly constrained by flavor physics
\cite{Barbier:2004ez}, and thus they are less relevant for the LSP
decay. However, the relative importance of $\lambda_{231}$ and the
$\lambda'_{jk}$ couplings cannot be determined. Table \ref{bounds}
shows the relevant experimental bounds for our model, as obtained in
references \cite{Barbier:2004ez,Kao:2009fg}. However, we cannot know
\emph{a priori} the relative importance of the different couplings and
general predictions concerning the LSP decay cannot be made (apart
from the aforementioned trilinear dominance).

\begin{table}[!tb]
\centering
\begin{tabular}{c c c}
\hline
Coupling & Bound & The bound comes from \\
\hline
$\lambda_{122}$ & $2.7 \cdot 10^{-2}$ & Neutrino masses \cite{Barbier:2004ez} \\
$\lambda_{133}$ & $1.6 \cdot 10^{-3}$ & Neutrino masses \cite{Barbier:2004ez} \\
$\lambda_{231}$ & 0.05 & Leptonic $\tau$ decay \cite{Kao:2009fg} \\
$\lambda'_{jk}$ ($j \ne k$) & $0.02 \,  - 0.47 \,$ & Several processes \cite{Barbier:2004ez} \\
$\lambda'_{jk}$ ($j = k$) & $3.3 \cdot 10^{-4} \, - 0.02 \,$ & $\beta \beta 0 \nu$ and $m_\nu$ \cite{Barbier:2004ez} \\
\hline
\end{tabular}
\caption{Experimental bounds for trilinear R-parity violating couplings. These limits were obtained by setting all SUSY masses to $100$ GeV. For the $\lambda'$ couplings the extreme cases are shown, with the most and least constrained couplings for each case. For more details see references \cite{Barbier:2004ez,Kao:2009fg}.}
\label{bounds}
\end{table}

Nevertheless, we emphasize that independent tests of the model and its
underlying flavor structure should be performed. In addition to the
aforementioned correlations among neutrino mixing angles,
unfortunately present in many flavor models, collider tests are
fundamental in order to disentangle the dynamics behind the observed
flavor pattern in neutrino mixing. Reference \cite{Bomark:2011ye}
addresses this issue by studying how one may discriminate between
different \rpv flavor operators leading to the decays of a
neutralino LSP at the LHC. Similar works exist in the case of b-\rpv
\cite{Mukhopadhyaya:1998xj,Hirsch:2000ef,Porod:2000hv,Hirsch:2002ys,Thomas:2011kt,deCampos:2012pf},
where one can find a strong correlation between LSP decays and
neutrino mixing angles. Finally, it is also possible to distinguish
between bilinear and trilinear violation of R-parity if the LSP is a
slepton. In that case, its decays properties can be used to determine
the relative importance of the \rpv couplings, as shown in
\cite{Bartl:2003uq}.

\section{Conclusions}
\label{sec:conclusions}

We studied a simple extension of the MSSM assuming an abelian $Z_3$
flavor symmetry.  Neutrino masses arise at 1-loop from bilinear and
trilinear R-parity breaking operators. If the soft SUSY breaking terms
were flavor blind, one would obtain a $\mu-\tau$ invariant neutrino
mass structure, thus implying a vanishing reactor angle. However, this
is cured by renormalization group running effects, naturally present
in supersymmetric theories. We have shown how this model can
accommodate a neutrino mixing pattern in accordance to data. In
particular, we have studied the correlations between mixing angles
that allow us to link different deviations from tri-bimaximal
mixing. The model also predicts large values for the lightest neutrino
mass and the effective $m_{ee}$ $0\nu 2\beta$ parameter, both at the
reach of current and future experiments.  Finally, since the LSP
decays one in principle has a rich phenomenology at the LHC. Due to
the structure of the model, we predict a clear dominance of the
trilinear couplings in the LSP decay.

\section*{Acknowledgements}

We thank M. Hirsch, S. Morisi and J.F.W. Valle for useful discussions.
Work supported by the Spanish MEC under grants FPA2008-00319/FPA,
FPA2011-22975 and MULTIDARK CSD2009-00064 (Consolider-Ingenio 2010
Programme), by Prometeo/2009/091 (Generalitat Valenciana). E. P. is
supported by CONACyT (Mexico). A.V. acknowledges support by the ANR
project CPV-LFV-LHC {NT09-508531}.

\end{document}